\documentclass[12pt]{iopart}
\usepackage{iopams}
\usepackage{subfigure}
\usepackage{graphicx}
\usepackage{epstopdf}
\usepackage{tikz}

\newcommand{\hi}{H_i} % initial Hamiltonian
\newcommand{\hf}{H_f} % final Hamiltonian
\newcommand{\mingap}{\Delta_{\mathrm{min}}} % minimum gap
\newcommand{\melement}{\mathcal{E}} % matrix element of dH/ds with instantaneous ground and excited states
\newcommand{\identity}{\mathbb{I}} % identity matrix
\newcommand{\Gaussian}[1]{\mathcal{N}(0, #1^2)} % Gaussian distribution (argument: standard deviation)
\newcommand{\uniform}[1]{\mathcal{U}(-#1, #1)} % uniform distribution (argument: positive cut-off)

\newcommand{\bra}[1]{\left\langle #1 \right|}
\newcommand{\ket}[1]{\left| #1 \right\rangle}
\newcommand\braket[3][]{
  \def\temp{#1}
  \ifx\temp\empty \left\langle #2 \left| #3 \right.\right\rangle
  \else \left\langle #2 \left| #1 \right| #3 \right\rangle
  \fi
}

\begin{document}
\title[Adiabatic quantum computing success probability]{Relationship between minimum gap and success probability in adiabatic quantum computing}
\author{M~Cullimore\footnote{Current address: Laser and Calibration Products Division, Renishaw plc, Bath Road, Woodchester GL5 5EY, United Kingdom},
M~J Everitt,
M~A~Ormerod\footnote{Current address: BAE Systems, West Hanningfield Road, Great Baddow, Chelmsford CM2 8HN, United Kingdom},
J~H~Samson,
R~D~Wilson and
A~M~Zagoskin} 
\address{Department of Physics, Loughborough University, Loughborough, Leicestershire LE11 3TU, United Kingdom}
 \ead{J.H.Samson@lboro.ac.uk}
\begin{abstract}
We explore the relationship between two figures of merit for an adiabatic quantum computation process: the success probability  $P$ and the minimum gap $\mingap$ between the ground and first excited states, investigating to what extent the success probability for an ensemble of problem Hamiltonians can be fitted by a function of $\mingap$ and the computation time $T$. We study a generic adiabatic algorithm and show that a rich structure exists in the distribution of $P$ and $\mingap$.  In the case of two qubits, $P$ is to a good approximation a function of $\mingap$, of the stage in the evolution at which the minimum occurs and of $T$.  This structure persists in examples of larger systems.
\end{abstract}
 % \submitto{\JPA \today} 

\maketitle

\section{Introduction}

The promise of a qualitative advantage of quantum computers over  classical ones in solving certain classes of problems has led to a massive effort in theoretical and experimental investigation of controlled, quantum-coherent systems. The standard circuit model (CM) of quantum computing is analogous to classical computing in the sense of requiring a sequence of logic gate operations. However, the requirement of precise time-dependent control of individual qubits in the quantum case is hard to achieve experimentally while still maintaining the quantum coherence of the system. A number of alternative approaches have been proposed, of which \emph{adiabatic quantum computing} (AQC) is a promising example. This involves the evolution of a quantum system from a simple Hamiltonian with an easily-prepared ground state to a Hamiltonian that encodes the problem to be solved, and whose ground state encodes the solution. If the system is  prepared in the initial ground state and the time evolution occurs slowly enough to satisfy the adiabatic theorem, the final state will have a large overlap with the ground state. Measurement in the computational basis will then yield the desired solution with high probability \cite{farhi-2000}.   

Several authors have demonstrated polynomial equivalence between AQC and the CM, mapping the latter onto an AQC with $3$-local interactions between qubits or $2$-local interactions between $6$-state qudits in two dimensions \cite{aharonov-2005}, or with $2$-local interactions between qubits on a two-dimensional lattice (but requiring two or more control Hamiltonians) \cite{mizel-2007,oliveira-2008}.

Despite these proofs of equivalence between AQC and CM, it is clear that there are classes of problems more suited to one or the other; in addition, AQC is believed to be more robust against decoherence \cite{childs-2001}, although the effects of decoherence \cite{sarandy-2005} and noise \cite{roland-2005,ashhab-2006} imply an optimal computation time, beyond which errors increase.    The type of problems most suited to AQC include optimization problems, where the requirement is to find the global minimum of a cost function $f: \{0,1\}^{n}\rightarrow \mathbb R$, and the related decision problems, where the requirement is to demonstrate the existence of a good solution $y$ obeying $f_{y}<F$ for some specified $F$. Thus the existence proof of a polynomial-time AQC implementation of Shor's prime factorization algorithm does not help practical implementation: one rather starts afresh and maps factorization onto an optimization problem, as in the recent NMR factorization of $143$ \cite{xu-2012}.  Simulations of the travelling salesman problem show faster decay of residual energy (i.~e., tour length) through AQC than through classical simulated annealing, although other classical algorithms are faster \cite{martonak-2004,das-2008}.    Applications have also been found in graph theory, most recently in the evaluation of Ramsey numbers \cite{gaitan-2012}. 

The task of AQC is to find the ground state of a Hamiltonian $\hf$; this Hamiltonian encodes the problem under consideration and its (unknown) eigenvalues  determine the cost function.  A Hamiltonian $H(s)$ interpolates between a simple initial Hamiltonian, $\hi$, at time $t=0$ and the desired final Hamiltonian, $\hf$,  at the end of the computation $t=T$. Many interpolation schemes have been considered, which may optimize final-state fidelity but require some knowledge of the energy-level structure \cite{roland-2002,avron-2010} or  phase cancellation  \cite{wiebe-2012}. We therefore restrict consideration to the simple linear interpolation
\begin{equation}
H(s) = (1 - s)\hi + s\hf,
\label{eq:AQCH}
\end{equation}	
where $s\in[0,1]$ is the reduced time $s=t/T$.  The  eigenvalues and eigenstates of the Hamiltonian $H(s)$ of an $n$-qubit system are given by 
\begin{equation}\label{eq:eigen}
H(s)\ket{m; s}= E_m(s)\ket{m; s},\;\mbox{with}\; E_0(s) \leqslant E_1(s) \leqslant \cdots \leqslant E_{2^n - 1}(s).
\end{equation}
The instantaneous state of the system is given by $\ket{\psi(s)}$, the solution of Schr\"odinger's equation, which in reduced time (and $\hbar=1$) reads
\begin{equation}
\frac \rmd {\rmd s}\ket{\psi(s)} = -\rmi TH(s)\ket{\psi(s)}.
\end{equation}
The system is prepared in the (non-degenerate) ground state  of $\hi$: $\ket{\psi(0)} = \ket{0;0}$.

At the end of the evolution a suitable figure of merit is the closeness of the state vector, $\ket{\psi(1)}$,  to the desired result, $\ket{0; 1}$. This is provided by the \emph{success probability}
\begin{equation}
\label{eq:P}
P_{n}(\hi,\hf,T) = \bigl|\braket{0; 1}{\psi(1)}\bigr|^2.
\end{equation}
The subscript $n$, denoting the number of qubits, will be omitted except where a distinction needs to be made.

In practical optimization problems, a low-cost solution that is not necessarily the global optimum often suffices.  Here the \emph{energy error}
\begin{equation}
\Delta E = \langle\psi(1)|\hf|\psi(1)\rangle- \langle0;1|\hf|0;1\rangle
\label{eq:eerror}
\end{equation}
is a suitable figure of merit. Approximate Adiabatic Quantum Computing (AAQC) aims to reduce this error \cite{zagoskin-2007}.  For some purposes other characterizations of the final-state distribution $P_{m}=\bigl|\braket{m; 1}{\psi(1)}\bigr|^2$ may be more appropriate figures of merit.  In the present work we shall concentrate on the success probability.

We require $2^{n}$ parameters to specify the Hamiltonian $\hf$. One of the aims of this work is to investigate to what extent the success probability (\ref{eq:P}) can be approximated as a function $P(\hi,\hf,T) \approx \tilde P(\{\alpha_{j}\},T)$, where $\{\alpha_{j}(\hi,\hf),j=1\ldots M\}$ is a  small ($n$-independent) number of parameters characterizing the initial and final Hamiltonians.  The most important dependence is expected to be on the \emph{minimum gap} between ground state and first excited state
\begin{equation}
\label{eq:mingap}
\mingap = \min_{0 \leqslant s \leqslant 1} \bigl(E_1(s) - E_0(s)\bigr)
\end{equation}
which occurs at the reduced time(s) $s=s^{*}$:
\begin{equation}
\label{eq:sstar}
E_{1}(s^{*})-E_{0}(s^{*})=\mingap.
\end{equation}

While it has long been known \cite{born-1928,messiah-1961} that this probability tends to unity for slow evolution: \begin{equation}
\label{eq:lim}
\lim_{T\rightarrow\infty}P(\hi,\hf,T)=1 \mbox{  if  } \mingap>0,
\end{equation} the precise statement of this \emph{adiabatic theorem} has been the subject of much debate in recent years \cite{farhi-2001,sarandy-2004,marzlin-2004,tong-2005,wu-2005,jansen-2007,amin-2009,shevchenko-2010}.  The original statement in the context of AQC \cite{farhi-2000}  was that the adiabatic condition
\begin{equation}
T \gg \frac{\melement}{\mingap^2},
\label{eq:adi1}
\end{equation}
where
\begin{equation}
\label{eq:melement}
\melement = \max_{0 \leqslant s \leqslant 1}\left|\braket[\frac{\rmd H}{\rmd s}]{1; s}{0; s}\right|,
\end{equation}
guarantees  $P$ to be very close to $1$.  While this only considers transitions into the first excited state, such transitions will dominate in most situations. Sarandy \emph{et al} \cite{sarandy-2004} derived such a result, with $\melement$ taken as the maximum over all matrix elements to excited states. If $\melement$ is considered constant (of the order of a typical eigenvalue of $\hi$), $\mingap$ determines the required $T$. 

It is however sufficient (see, for example, Ref. \cite{amin-2009}) to require an evolution time
\begin{equation}
\label{eq:adi2}
T \gg \max_{0 \leqslant s \leqslant 1} \frac{ \left|\bra{m;s}\frac \rmd {\rmd s} \ket{0;s}\right|} {E_{m}(s)-E_{0}(s)}= \max_{0 \leqslant s \leqslant 1} \frac {\left|\bra{m;s}\frac {\rmd H} {\rmd s} \ket{0;s}\right|}{\left(E_{m}(s)-E_{0}(s)\right)^{2}},
\end{equation}
for all excited states $m>0$.  (In the present context we are restricting consideration to evolution of the ground  state.) Some authors \cite{marzlin-2004,tong-2005} have  claimed  counterexamples to the above criterion. However, these counterexamples include a resonant term, which is absent from our interpolating Hamiltonian (\ref{eq:AQCH}).

For practical purposes the knowledge that the success probability tends to unity in the infinite-time limit is of less interest than knowledge of parameters governing success for finite evolution times; it is this question that motivates our study.  The minimum gap $\mingap$ is usually considered to be the dominant parameter determining the success probability for a given evolution time. These two variables, $P$ and $\mingap$, are both used in the literature to quantify the performance of a given computation, and are assumed to increase monotonically with each other. The question of how either of these variables varies with system size is an important one that is often addressed. However, the exact nature of the correlation between these two important figures of merit has not been fully explored.  

We explore the relationship between $P$ and $\mingap$ by looking at the statistical distributions of these two variables over an ensemble of problem Hamiltonians ($\hf$) for fixed computation times $T$. We start by considering a simple two-qubit system and show that a rich structure arises in the scatter plots of success probability against $\mingap$.  We then go on to look at the scatter plots in three-, four- and five-qubit systems and find that, although some of the finer details of the structure are washed out, some remain.

\section{A generic adiabatic algorithm}

We wish to look at the distribution of the success probability and $\mingap$ over a large set of problem Hamiltonians. We use a simple, yet generic, model that is scalable and can be readily solved numerically. For $\hi$ we use  a transverse field of unit magnitude acting on all the qubits:
\begin{equation}
\hi
= -\sum_{i = 1}^n \sigma_x^{(i)}
\equiv -\sum_{i = 1}^n \underbrace{\identity \otimes \cdots \otimes \identity \otimes\sigma_x}_{i} \otimes \underbrace{\identity \otimes \cdots \otimes \identity}_{n-i},
\end{equation}
where $\sigma_{x},\sigma_{y},\sigma_{z}$ denote the usual Pauli matrices, $n$ is the number of qubits in the system; the $2^{n}\times 2^{n}$ matrix $\sigma_x^{(i)}$ acts on the $i$th qubit. The (non-degenerate) ground state of $\hi$ is an equal superposition
\begin{equation} \label{eq:gs}
\ket{0;0}=2^{-n/2}\sum_{y=0}^{2^{n}-1}\ket y
\end{equation}
of all $2^n$ computational basis states.

For $\hf$, we use a random-energy Hamiltonian, diagonal in the computational basis, where all $z$-axis couplings between the $n$ qubits are realized:
\begin{equation}
\hf
= \sum_{x = 1}^{2^n - 1} J_x \bigotimes_{i = 1}^n (\sigma_z)^{x_i}
= \sum_{y = 0}^{2^n - 1} f_y \ket{y} \! \bra{y}.
\end{equation}
Here $x_i$ is the $i$th digit in the binary representation of $x$. Where there are $k$ non-zero bits in the binary representation of $x$, the coupling constant $J_{x}$ represents a $k$-local interaction (a non-trivial interaction between $k$ qubits).  The $\{J_{x}\}$ will be selected from a suitable random distribution; we fix the trivial energy shift $J_{0}=0$. $\hf$ is diagonal in the computational basis so that the binary-ordered set of states $\ket y$ is a permutation of the energy-ordered set of states $\ket{m;1}$ defined in Eq.~\ref{eq:eigen} (in the generic case where the latter are non-degenerate). A Hamiltonian of this type can  be used to encode any finite computational optimization problem (minimization of a function $f: \{0,1\}^{n}\rightarrow \mathbb R$) by choice of the $\{J_x\}$. It is important to note that only $1$- and $2$-local interactions are experimentally feasible; however, higher-order interactions may be reducible to such terms at the cost of auxiliary qubits \cite{mizel-2007,oliveira-2008,duan-2011,kempe-2006}.

For each sample in the scatter plots, we solve the Schr\"odinger equation numerically over the reduced time range $0 \le s \le 1$ for a given computation time, $T$, using the Dormand-Prince method \cite{dormand-1980}. This is an adaptive step-size algorithm; solutions accurate to fourth- and fifth-order in the step size $\Delta s$ are used to estimate the local error in the former. If it is less than the desired tolerance, then the fifth-order solution is used for the integration. Otherwise  $\Delta s$ is decreased.

\begin{figure}[!tb]
\begin{center}
\begin{tikzpicture}[scale=1.0]
\pgftext[bottom,left]{%
\includegraphics[width=0.7\textwidth, trim = 15mm 10mm 0mm 0mm,clip]{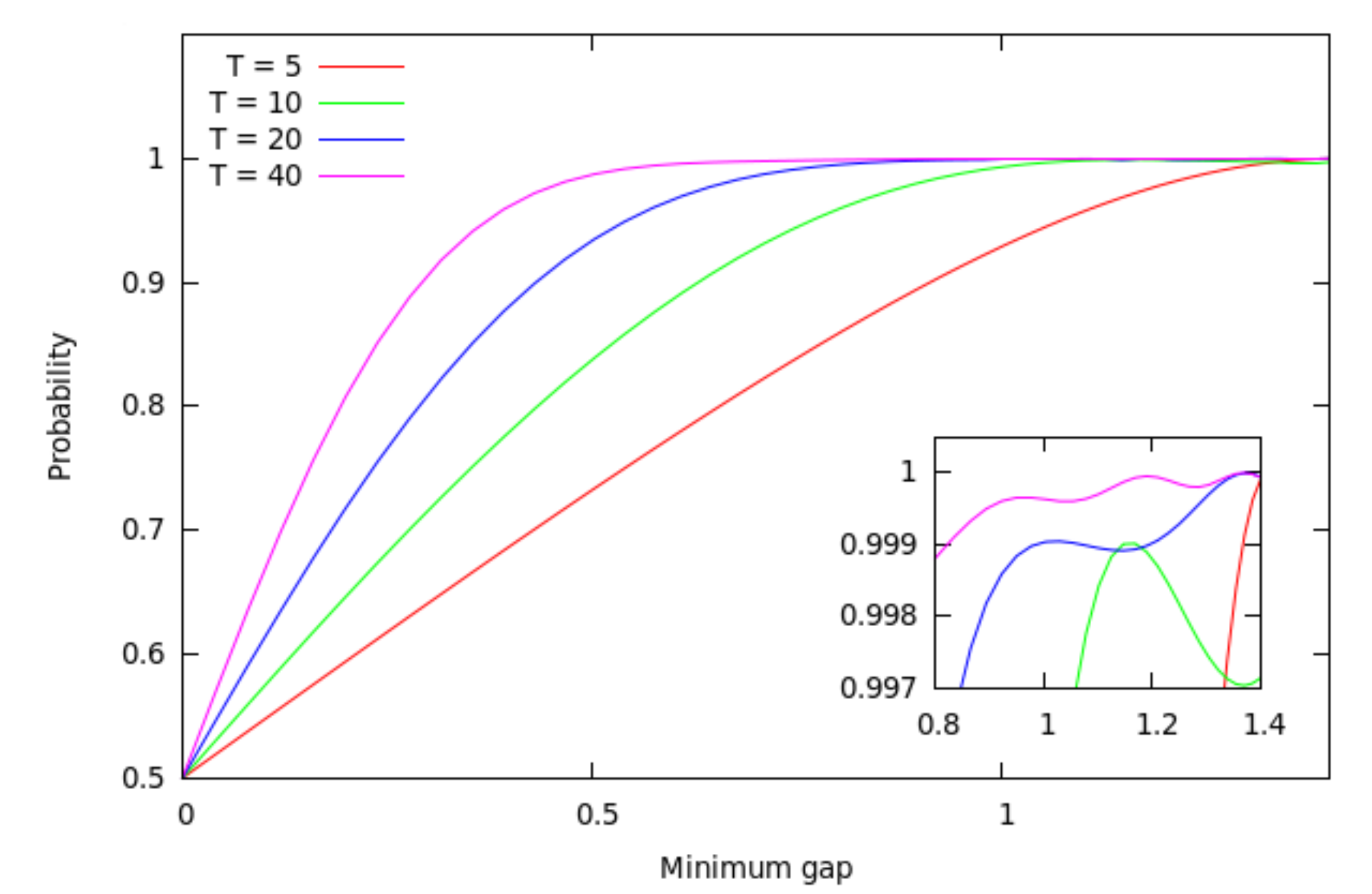}%
}%
\node (abscissa) at (6,-0.5) {\small Minimum gap ($\mingap$)};
\node (coodinate) [rotate=90] at (-0.2,3.5) {\small Probability ($P$)};
\end{tikzpicture}
\end{center} 
\caption{(Colour online) One-qubit success probability  for $T = 5, 10, 20, 40$ (bottom to top), plotted for $0\le J_{1}\le 1$. The magnified section indicates that the probability is not monotonic in either $\mingap$ or $T$.}
\label{fig:P1}
\end{figure}

For comparison with later scatter plots, figure~\ref{fig:P1} plots the probability $P(J_{1},T)$ against minimum gap 
\begin{equation}
\mingap = \frac {2\left|J_{1}\right|}{\sqrt{1+J_{1}^{2}}}
\end{equation}
 for a single qubit.  Since the final Hamiltonian is specified by a single parameter, $P$ is a (not quite monotonic) function of $\mingap$ and $T$.  A test of accuracy of the simulation is that the small component of the final state should be real:  $\mathrm{Im} \braket{1; 1}{\psi(1)}=0$ (which can be verified analytically).  Our numerical calculations reproduce this to high precision.
\section{Two-qubit simulations}
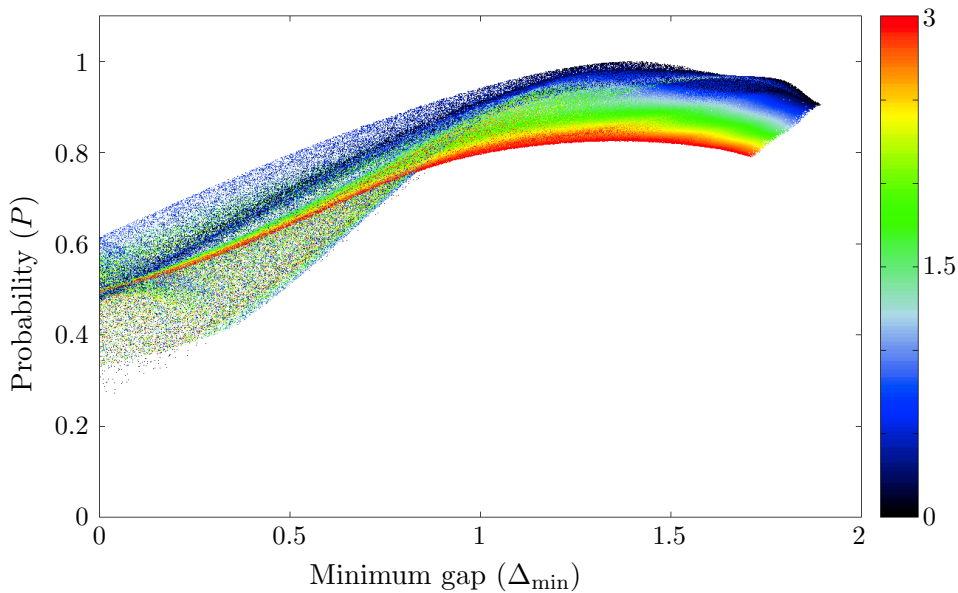
\begin{figure}[!tb]
\begin{center}
\begin{tikzpicture}[scale=1.0]
\pgftext[bottom,left]{%
\includegraphics[width=0.7\textwidth, trim = 40mm 20mm 40mm 0mm,clip]{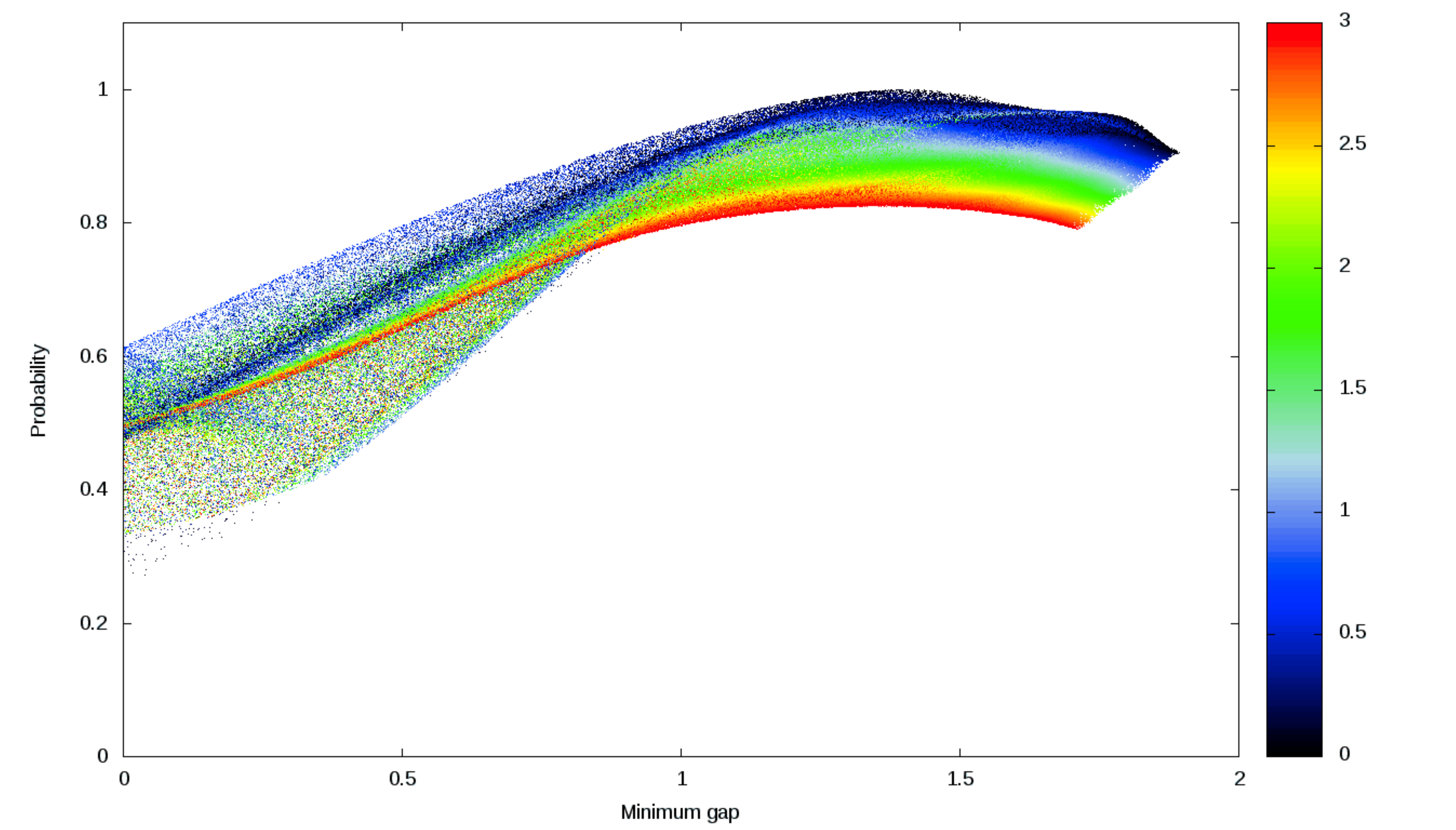}%
}%
\node (abscissa) at (4.6,-0.7) {\small Minimum gap ($\mingap$)};
\node (coodinate) [rotate=90] at (-1,3.075) {\small Probability ($P$)};

\node (0a) at (0.0,-0.15) {\footnotesize $0$};
\node (1a) at (2.525,-0.15) {\footnotesize $0.5$};
\node (2a) at (5.05,-0.15) {\footnotesize $1$};
\node (3a) at (7.575,-0.15) {\footnotesize $1.5$};
\node (4a) at (10.1,-0.15) {\footnotesize $2$};

\node (0c) [left] at (0,0.1) {\footnotesize $0$};
\node (0c) [left] at (0,1.31) {\footnotesize $0.2$};
\node (0c) [left] at (0,2.52) {\footnotesize $0.4$};
\node (0c) [left] at (0,3.73) {\footnotesize $0.6$};
\node (0c) [left] at (0,4.94) {\footnotesize $0.8$};
\node (0c) [left] at (0,6.15) {\footnotesize $1$};

\node (0b) [right] at (10.8,0.1) {\footnotesize $0$};
\node (0b) [right] at (10.8,3.45) {\footnotesize $1.5$};
\node (0b) [right] at (10.8,6.75) {\footnotesize $3$};

\end{tikzpicture}
\end{center}
\caption{(Colour online) Success probability against minimum gap for a two-qubit system at a computation time of $T = 5$. The $J_i$ have been chosen from the uniform distribution $\uniform{3}$ for  $500,000$ random problem instances. The data points are coloured by $|J_3|$.}
\label{fig:2qb_uniform}
\end{figure}

For two or more qubits, the success probability will no longer collapse to a function $\tilde P(\mingap,T)$ of the minimum gap and computation time.   Figure \ref{fig:2qb_uniform} is a scatter plot of success probability $P(\{J_{x}\},T)$ against minimum gap $\mingap(\{J_{x}\})$ for a large set of two-qubit problem instances, with the coupling coefficients $J_{1}\ldots J_{3}$ drawn from the uniform distribution $\uniform{3}$, and a short computation time $T = 5$. Observe the sharp upper and lower edges. The lower bound of the success probability is always $1/4$ for infinitesimally small $\mingap$. This arises when $J_1 = J_2 = J_3 = 0$: with a four-fold degeneracy at $s = 1$  the system remains in its original ground state (\ref{eq:gs}).  

\begin{figure}[hbt]
\begin{center}

\subfigure{
\begin{tikzpicture}[scale=0.8]
\pgftext[bottom,left]{%
\includegraphics[width=0.45\textwidth, trim = 40mm 20mm 47mm 0mm,clip]{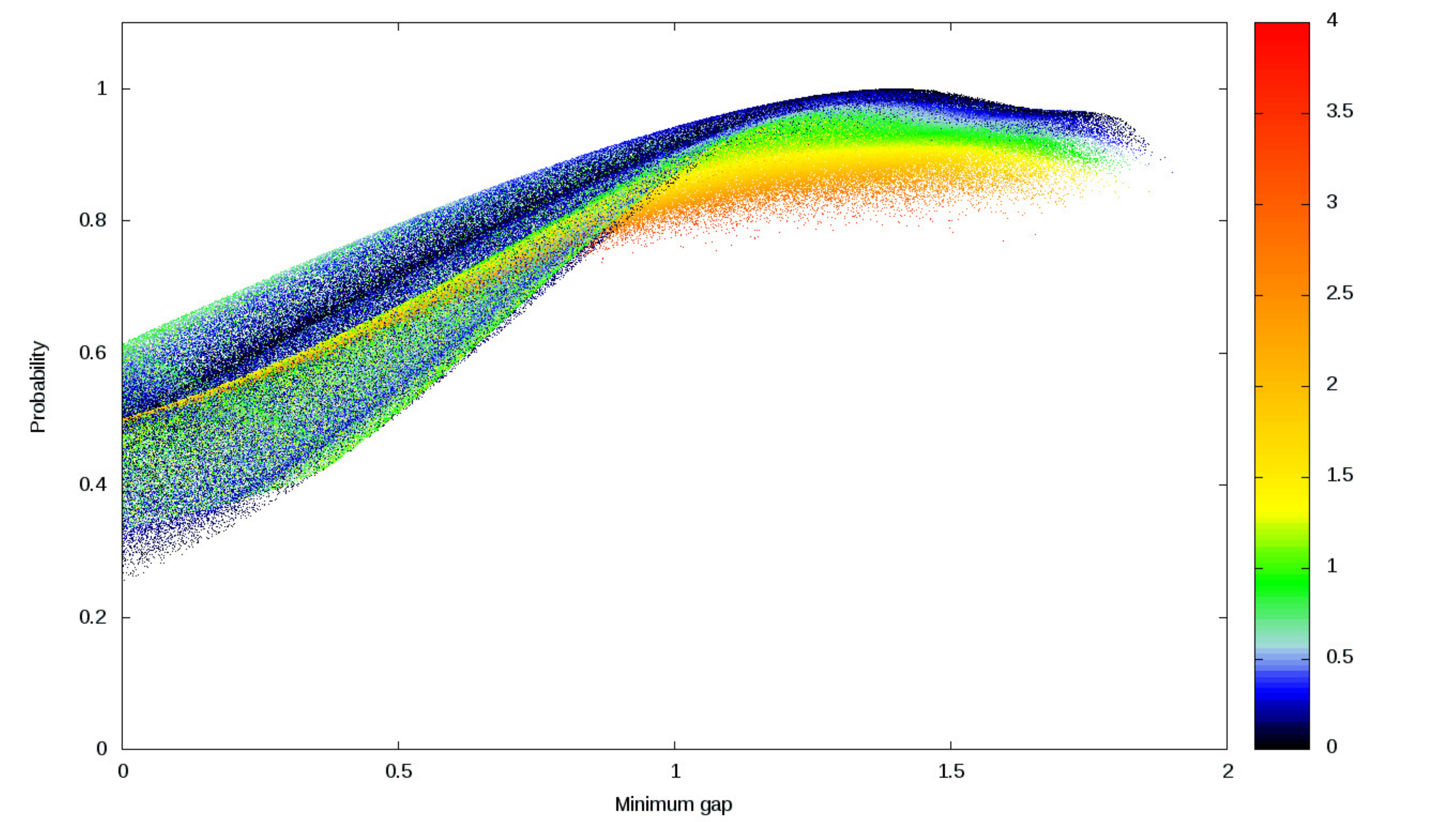}%
}%
\node (abscissa) at (3.25,-0.7) {\footnotesize Minimum gap ($\mingap$)};
\node (coodinate) [rotate=90] at (-1,2) {\footnotesize Probability ($P$)};
\node (figLabel) at (5.5,0.75) {\footnotesize $T = 5$};

\node (0a) at (0.0,-0.15) {\scriptsize $0$};
\node (1a) at (1.625,-0.15) {\scriptsize $0.5$};
\node (2a) at (3.25,-0.15) {\scriptsize $1$};
\node (3a) at (4.875,-0.15) {\scriptsize $1.5$};
\node (4a) at (6.5,-0.15) {\scriptsize $2$};

\node (0c) [left] at (0,0.1) {\scriptsize $0$};
\node (0c) [left] at (0,0.88) {\scriptsize $0.2$};
\node (0c) [left] at (0,1.66) {\scriptsize $0.4$};
\node (0c) [left] at (0,2.44) {\scriptsize $0.6$};
\node (0c) [left] at (0,3.22) {\scriptsize $0.8$};
\node (0c) [left] at (0,4) {\scriptsize $1$};

\node (0b) [right] at (7,0.1) {\scriptsize $0$};
\node (0b) [right] at (7,2.2) {\scriptsize $2$};
\node (0b) [right] at (7,4.4) {\scriptsize $4$};
\end{tikzpicture}} \\

\subfigure{
\begin{tikzpicture}[scale=0.8]
\pgftext[bottom,left]{%
\includegraphics[width=0.45\textwidth, trim = 40mm 20mm 50mm 0mm,clip]{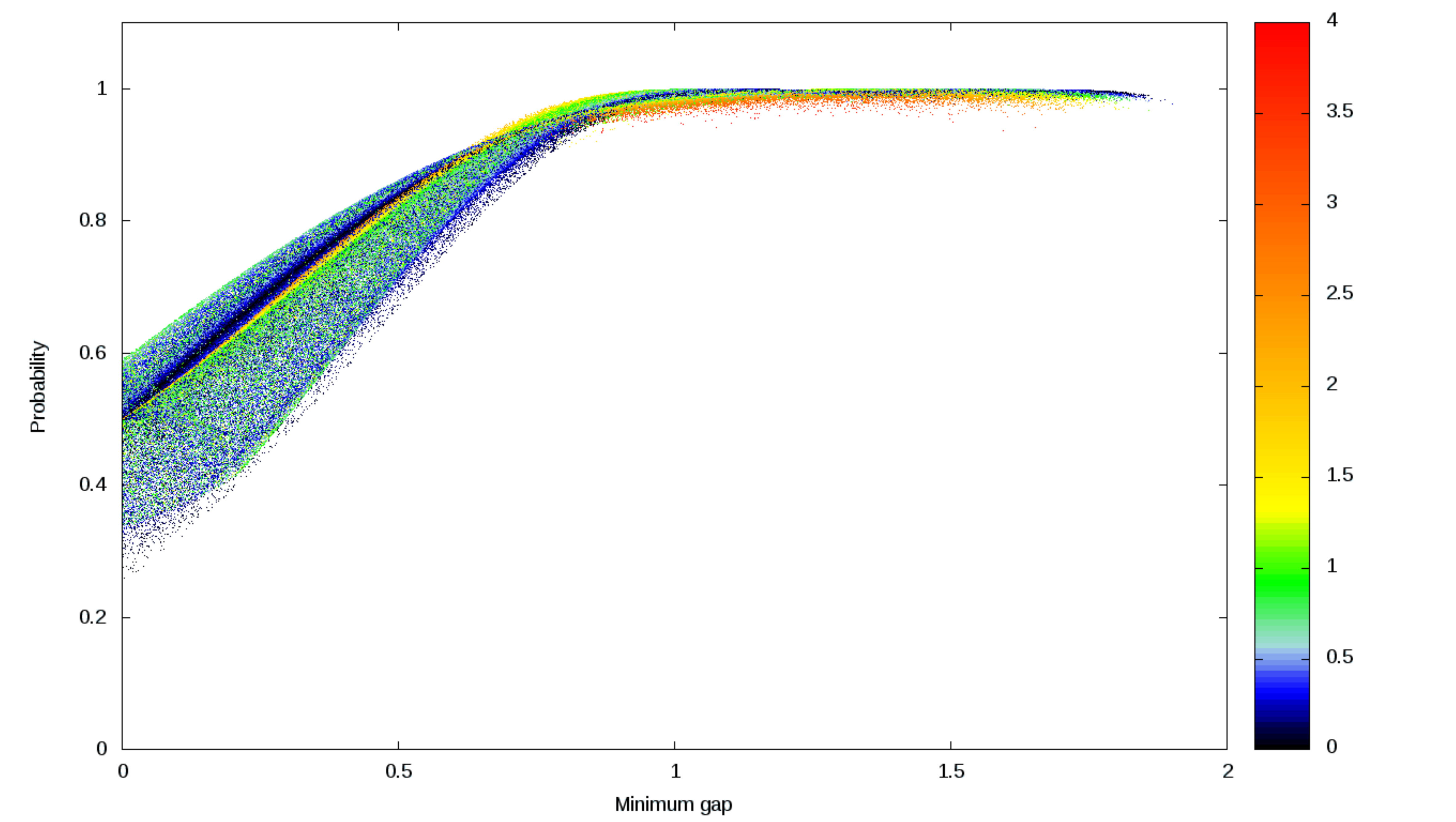}%
}%
\node (abscissa) at (3.25,-0.7) {\footnotesize Minimum gap ($\mingap$)};
\node (coodinate) [rotate=90] at (-1,2) {\footnotesize Probability ($P$)};
\node (figLabel) at (5.5,0.75) {\footnotesize $T = 5$};

\node (0a) at (0.0,-0.15) {\scriptsize $0$};
\node (1a) at (1.625,-0.15) {\scriptsize $0.5$};
\node (2a) at (3.25,-0.15) {\scriptsize $1$};
\node (3a) at (4.875,-0.15) {\scriptsize $1.5$};
\node (4a) at (6.5,-0.15) {\scriptsize $2$};

\node (0c) [left] at (0,0.1) {\scriptsize $0$};
\node (0c) [left] at (0,0.88) {\scriptsize $0.2$};
\node (0c) [left] at (0,1.66) {\scriptsize $0.4$};
\node (0c) [left] at (0,2.44) {\scriptsize $0.6$};
\node (0c) [left] at (0,3.22) {\scriptsize $0.8$};
\node (0c) [left] at (0,4) {\scriptsize $1$};

\node (0b) [right] at (7,0.1) {\scriptsize $0$};
\node (0b) [right] at (7,2.2) {\scriptsize $2$};
\node (0b) [right] at (7,4.4) {\scriptsize $4$};
\end{tikzpicture}} \\

\subfigure{
\begin{tikzpicture}[scale=0.8]
\pgftext[bottom,left]{%
\includegraphics[width=0.45\textwidth, trim = 40mm 20mm 50mm 0mm,clip]{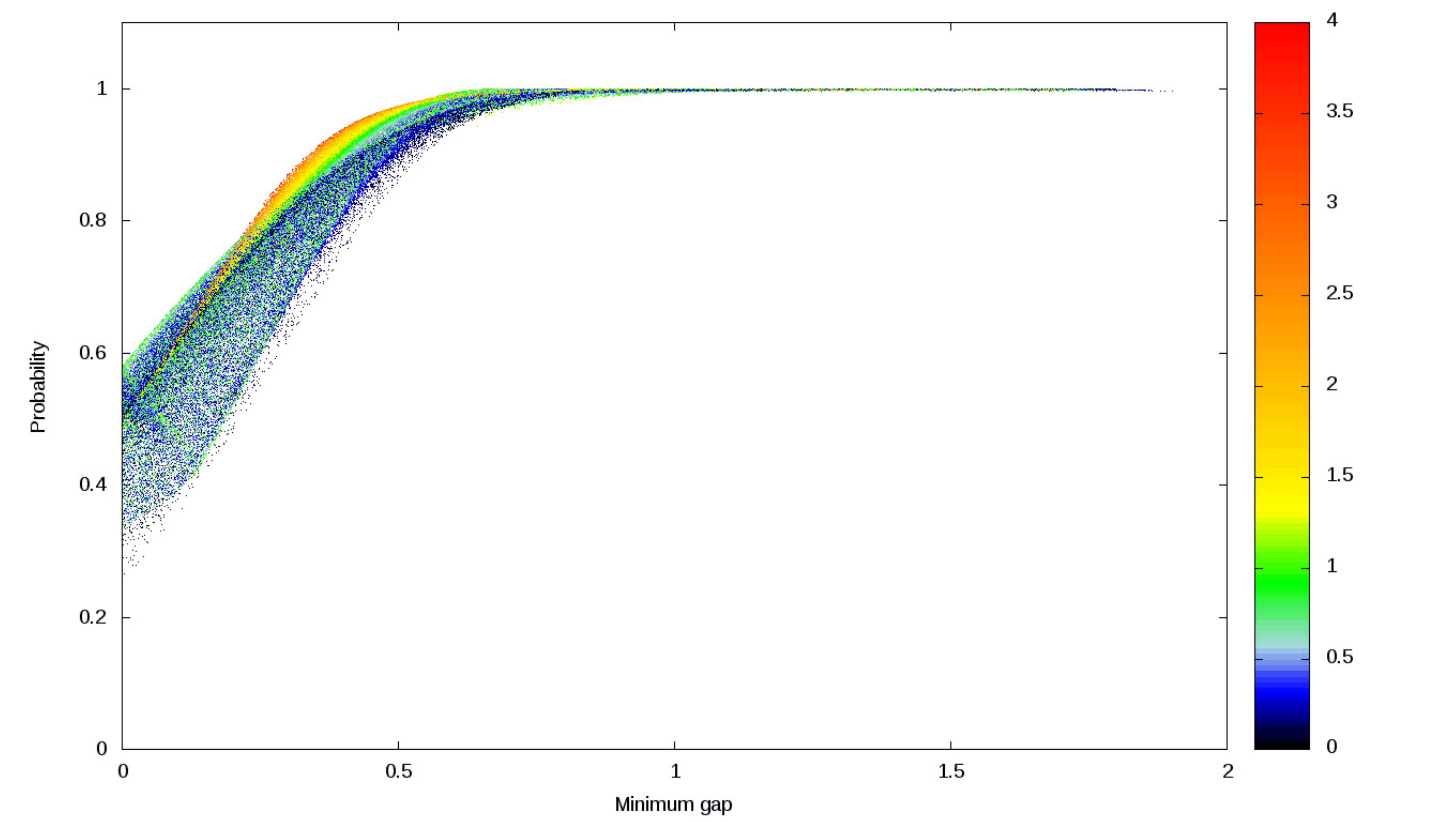}%
}%
\node (abscissa) at (3.25,-0.7) {\footnotesize Minimum gap ($\mingap$)};
\node (coodinate) [rotate=90] at (-1,2) {\footnotesize Probability ($P$)};
\node (figLabel) at (5.5,0.75) {\footnotesize $T = 5$};

\node (0a) at (0.0,-0.15) {\scriptsize $0$};
\node (1a) at (1.625,-0.15) {\scriptsize $0.5$};
\node (2a) at (3.25,-0.15) {\scriptsize $1$};
\node (3a) at (4.875,-0.15) {\scriptsize $1.5$};
\node (4a) at (6.5,-0.15) {\scriptsize $2$};

\node (0c) [left] at (0,0.1) {\scriptsize $0$};
\node (0c) [left] at (0,0.88) {\scriptsize $0.2$};
\node (0c) [left] at (0,1.66) {\scriptsize $0.4$};
\node (0c) [left] at (0,2.44) {\scriptsize $0.6$};
\node (0c) [left] at (0,3.22) {\scriptsize $0.8$};
\node (0c) [left] at (0,4) {\scriptsize $1$};

\node (0b) [right] at (7,0.1) {\scriptsize $0$};
\node (0b) [right] at (7,2.2) {\scriptsize $2$};
\node (0b) [right] at (7,4.4) {\scriptsize $4$};
\end{tikzpicture}} \\

\subfigure{
\begin{tikzpicture}[scale=0.8]
\pgftext[bottom,left]{%
\includegraphics[width=0.45\textwidth, trim = 40mm 20mm 50mm 0mm,clip]{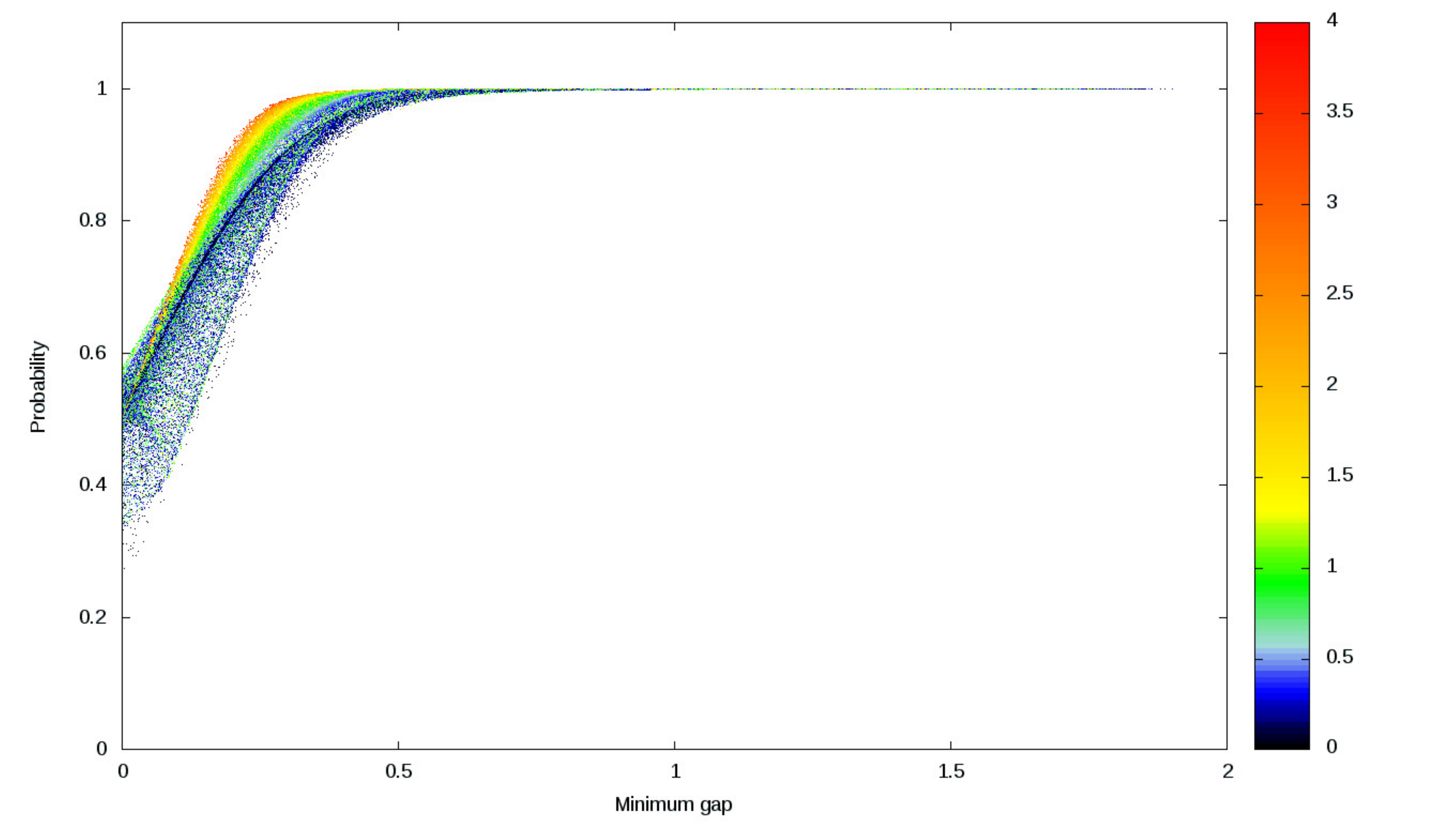}%
}%
\node (abscissa) at (3.25,-0.7) {\footnotesize Minimum gap ($\mingap$)};
\node (coodinate) [rotate=90] at (-1,2) {\footnotesize Probability ($P$)};
\node (figLabel) at (5.5,0.75) {\footnotesize $T = 5$};

\node (0a) at (0.0,-0.15) {\scriptsize $0$};
\node (1a) at (1.625,-0.15) {\scriptsize $0.5$};
\node (2a) at (3.25,-0.15) {\scriptsize $1$};
\node (3a) at (4.875,-0.15) {\scriptsize $1.5$};
\node (4a) at (6.5,-0.15) {\scriptsize $2$};

\node (0c) [left] at (0,0.1) {\scriptsize $0$};
\node (0c) [left] at (0,0.88) {\scriptsize $0.2$};
\node (0c) [left] at (0,1.66) {\scriptsize $0.4$};
\node (0c) [left] at (0,2.44) {\scriptsize $0.6$};
\node (0c) [left] at (0,3.22) {\scriptsize $0.8$};
\node (0c) [left] at (0,4) {\scriptsize $1$};

\node (0b) [right] at (7,0.1) {\scriptsize $0$};
\node (0b) [right] at (7,2.2) {\scriptsize $2$};
\node (0b) [right] at (7,4.4) {\scriptsize $4$};
\end{tikzpicture}}
\end{center}
\caption{(Colour online) Success probability against minimum gap for a two-qubit system at computation times of $T = 5$, $10$, $20$ and $40$. The $J_i$ have been chosen from the Gaussian distribution $\Gaussian{1}$ for the $500,000$ random problem instances. The data points are coloured by $|J_3|$.}
\label{fig:2qb_gaussian}
\end{figure}
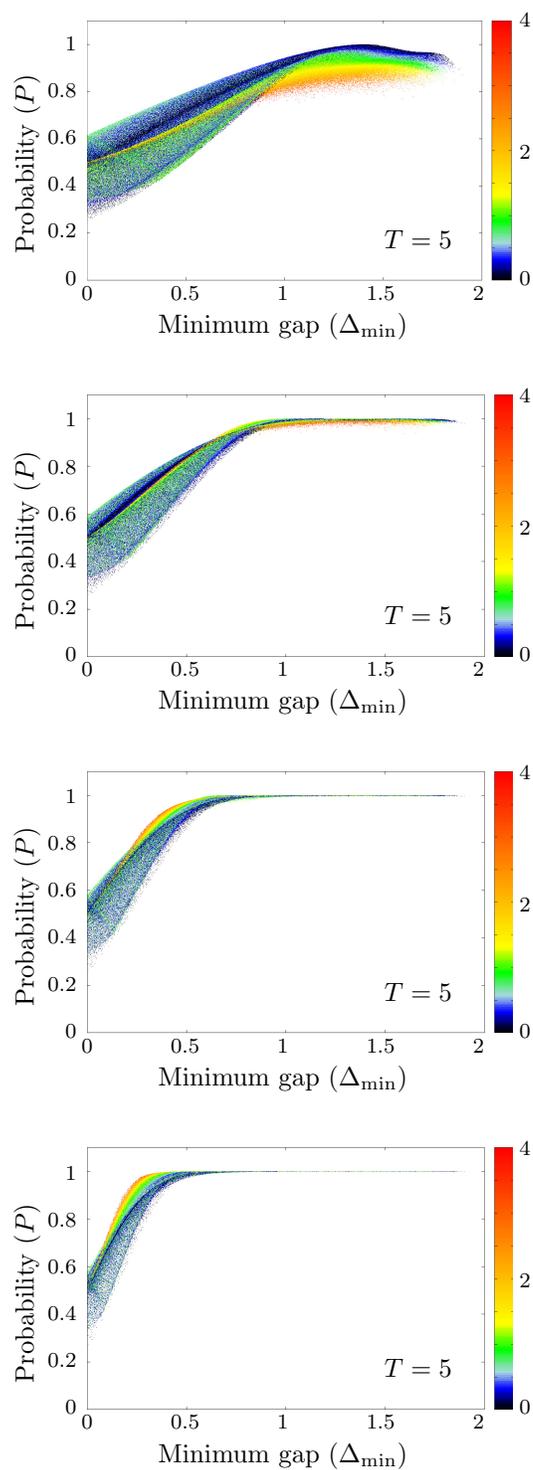

It is important to verify that this structure is independent of our choice of random distribution of coupling constants and that it is also not an artefact of the pseudo-random number generators used. Figure \ref{fig:2qb_gaussian} also shows scatter plots of success probability and minimum gap, but in this case the coupling coefficients are drawn from a Gaussian distribution, $\Gaussian{1}$ (mean $0$, standard deviation $1$). The trends and structure in the distributions are similar to those shown in figure \ref{fig:2qb_uniform}. However, there are some subtle differences in sharpness between the Gaussian and uniform cases. For a large minimum gap, the lowest probability occurs for large $|J_3|$, so we see a sharp cutoff in the uniform case but not in the Gaussian case. In general though, this shows that the results are independent of our choice of coupling constants and, as a different pseudo-random number generator routine was used, we can say that the results are not a numerical artefact.

Four computation times are shown: $T = 5$, $10$, $20$ and $40$. As $T$ increases, the distribution shifts and tends towards a success probability of $\lim_{T\rightarrow\infty}P=1$ for any $\mingap>0$, in agreement with the adiabatic theorem.

The two interesting features of these scatter plots are the well-defined sharp edges and the densely-populated bands. We colour the data points according to the strength of the two-qubit interactions, as this is a special direction in the two-qubit parameter space, which will determine the amount of entanglement during the evolution. It is clear that the bands correspond to groups of Hamiltonians with similar $|J_3|$. The bands where $J_3 = 0$ can be seen as two separable one-qubit evolutions for $J_1$ and $J_2$, so the total success probability $P_{2}$ is simply the product $P_{1}(J_{1},T)P_{1}(J_{2},T)$ of the one-qubit success probabilities shown in figure~\ref{fig:P1}:
\begin{equation}
\left(\tilde P_{1}(\mingap,T)\right)^{2} \lesssim  \tilde P_{2}(\mingap,T) \le \tilde P_{1}(\mingap,T)\;\mbox{for}\; J_{3}=0
\end{equation}
where
\begin{equation}
\mingap=\min\left(\frac{2|J_{1}|}{\sqrt{1+J_{1}^{2}}},\frac{2|J_{2}|}{\sqrt{1+J_{2}^{2}}}\right).
\end{equation}
Another interesting point to note is that the bands of similar $\hf$ gradually reverse in order in the distribution as the computation time $T$ is changed.

\begin{figure}[htb]
\centering
\includegraphics[width=0.5\textwidth]{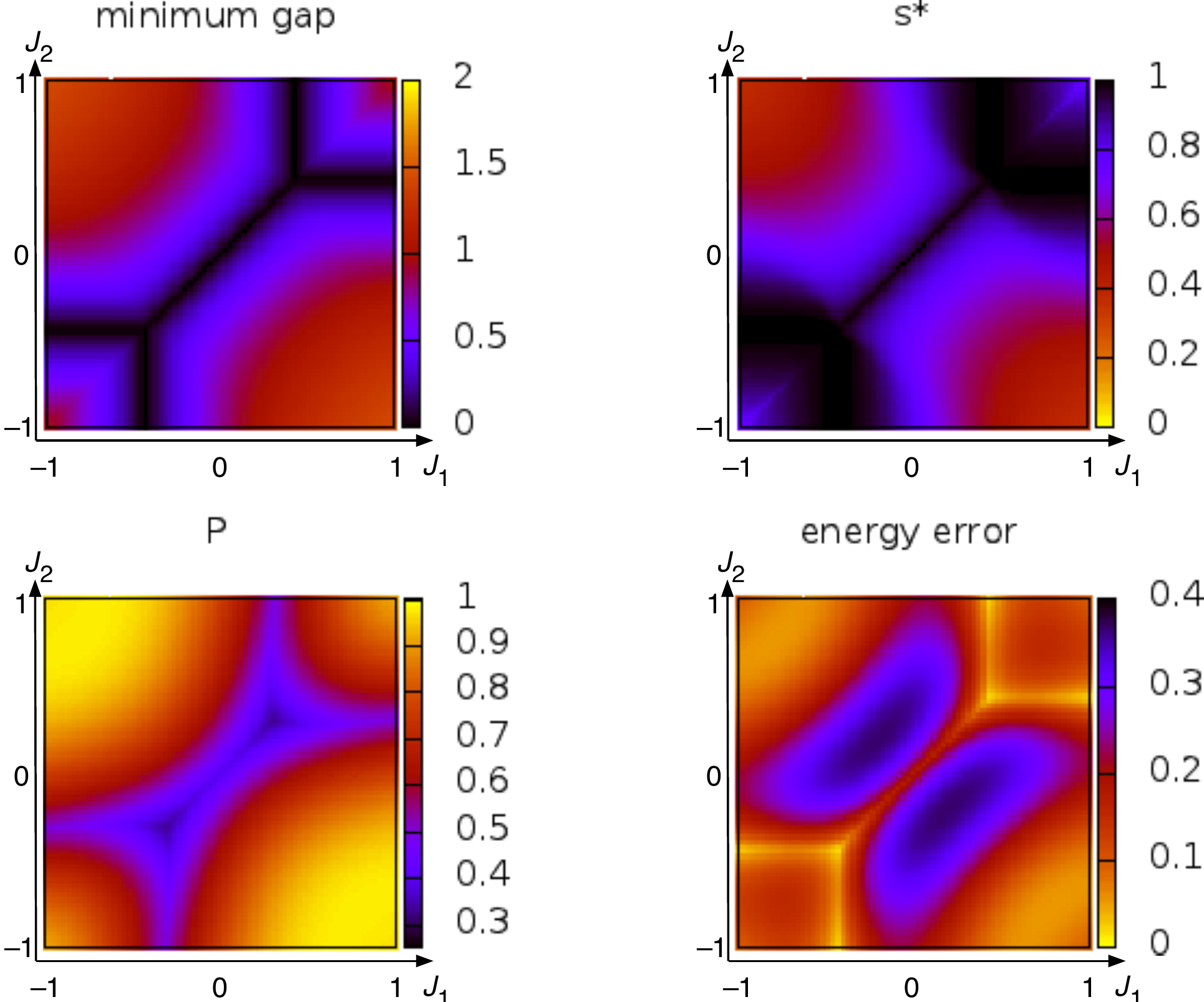}
\caption{(Colour online) Slice at $J_{3} = 0.43$  through ground-state phase diagram for an ensemble of two-qubit Hamiltonians.   Plots show  minimum gap $\mingap$ (top left), position of minimum gap $s^{*}$ (top right), success probability $P$  for $T = 5$  (bottom left) and energy error $\Delta E$  for $T = 5$ (bottom right).}
\label{fig:stability}
\end{figure}

We have supplemented the uniform random data with sets of $J_i$ chosen on a rectangular grid with the same cut-offs. These have the advantage that all problem Hamiltonians with a given value of $J_3$ can be plotted in the $(J_1, J_2)$-plane and coloured by their minimum gap or success probability; see figure \ref{fig:stability}. 

The energy structure in the case of two qubits can be simply characterized.  The final-state energies are given by
\begin{numparts}
\label{levels}
\begin{eqnarray}
f_{00} & = & J_{0} + J_{1} + J_{2} + J_{3},\\
f_{01} & = & J_{0} + J_{1} - J_{2} - J_{3},\\
f_{10} & = & J_{0} - J_{1} + J_{2} - J_{3},\\
f_{11} & = & J_{0} - J_{1} - J_{2} + J_{3}.
\end{eqnarray}
\end{numparts}

The ground-state phase diagram has tetrahedral symmetry $T_{d}$, with the regions of parameter space with ground states $\ket{00},\ket{01},\ket{10},\ket{11}$ separated by six planes meeting at the the four lines
\begin{numparts}
\label{boundary}
\begin{eqnarray}
J_{1}=J_{2}=J_{3} & >0 & \\
-J_{1}=J_{2}=J_{3} & <0 & \\
J_{1}=-J_{2}=J_{3} & <0 & \\
-J_{1}=-J_{2}=J_{3} &  >0 &
\end{eqnarray}
\end{numparts}
The eigenvalue dynamics has lower symmetry $D_{2d}$, since the degeneracy planes $f_{00}=f_{11}$ and $f_{01}=f_{10}$ admit entangled ground states and are inequivalent to the other four planes; this is borne out by the observation that neither the eigenvalue dynamics $E_{m}(s)$ nor the success probability is invariant under all permutations of the diagonal elements of $\hf$.  Identification of the symmetry structure of larger systems may cast further light on the $n$-qubit case. 

Figure \ref{fig:stability} shows a constant-$J_{3}$ slice through this phase diagram.  The degeneracy planes are clearly indicated in the minimum-gap plot; here the gap vanishes at $s^{*}=1$.  The lower two plots demonstrate the non-adiabaticity of the time evolution, with the success probability increasing (but not completely monotonically) with distance from the degeneracy planes.  The energy error is non-monotonic: it is small at the degeneracy planes (since the final state will have only a small admixture orthogonal to the degenerate ground states) and small where a large gap reduces the probability of transitions.

\begin{figure}[htb]
\centering
\includegraphics[width=0.755\textwidth]{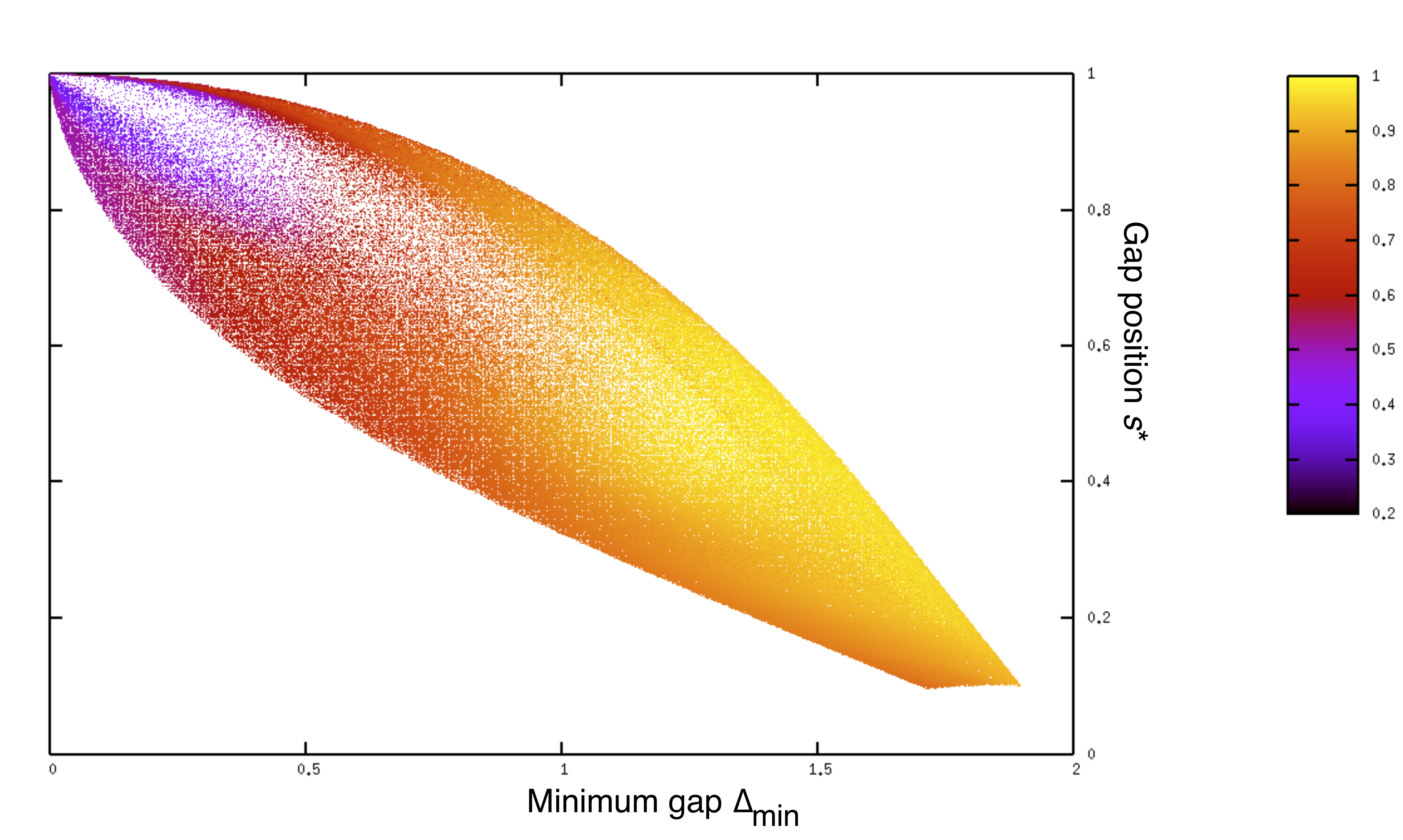}
\caption{(Colour online) Scatter plot of minimum gap position $s^{*}$   against minimum gap $\mingap$. The $J_i$ have been chosen from the uniform distribution $\uniform{3}$ for  $100,000$ random problem instances.   Points are coloured by the success probability $P$ at $T = 5$.}
\label{fig:3}
\end{figure}

These plots suggest a projection of a surface onto the $(\mingap,P)$ plane; we seek to find a suitable parameterization of the set of Hamiltonians to collapse it onto a low-dimensional surface.  We find that a plot of $P(\{J_{x}\},T)$ against the minimum gap $\mingap(\{J_{x}\})$ and the position $s^{*}(\{J_{x}\})$ of the gap indeed shows that all points lie close to a curved surface $\tilde P (\mingap,s^{*},T)$ (which rises with increasing $T$).  This is understandable, since those two parameters largely determine the shape of the lowest two energy levels.  Figure \ref{fig:3} shows a projection of this surface onto the $(\mingap,s^{*})$ plane. Visual inspection shows that the colour is to a good approximation a function only of position in the plane.   Note that the position of the points depends only on the Hamiltonian parameters, while the success probability depends also on the computation time. 

\begin{figure}[hbt]
\begin{center}
\subfigure{
\begin{tikzpicture}[scale=0.85]
\pgftext[bottom,left]{%
\includegraphics[width=0.45\textwidth, trim = 40mm 20mm 50mm 0mm,clip]{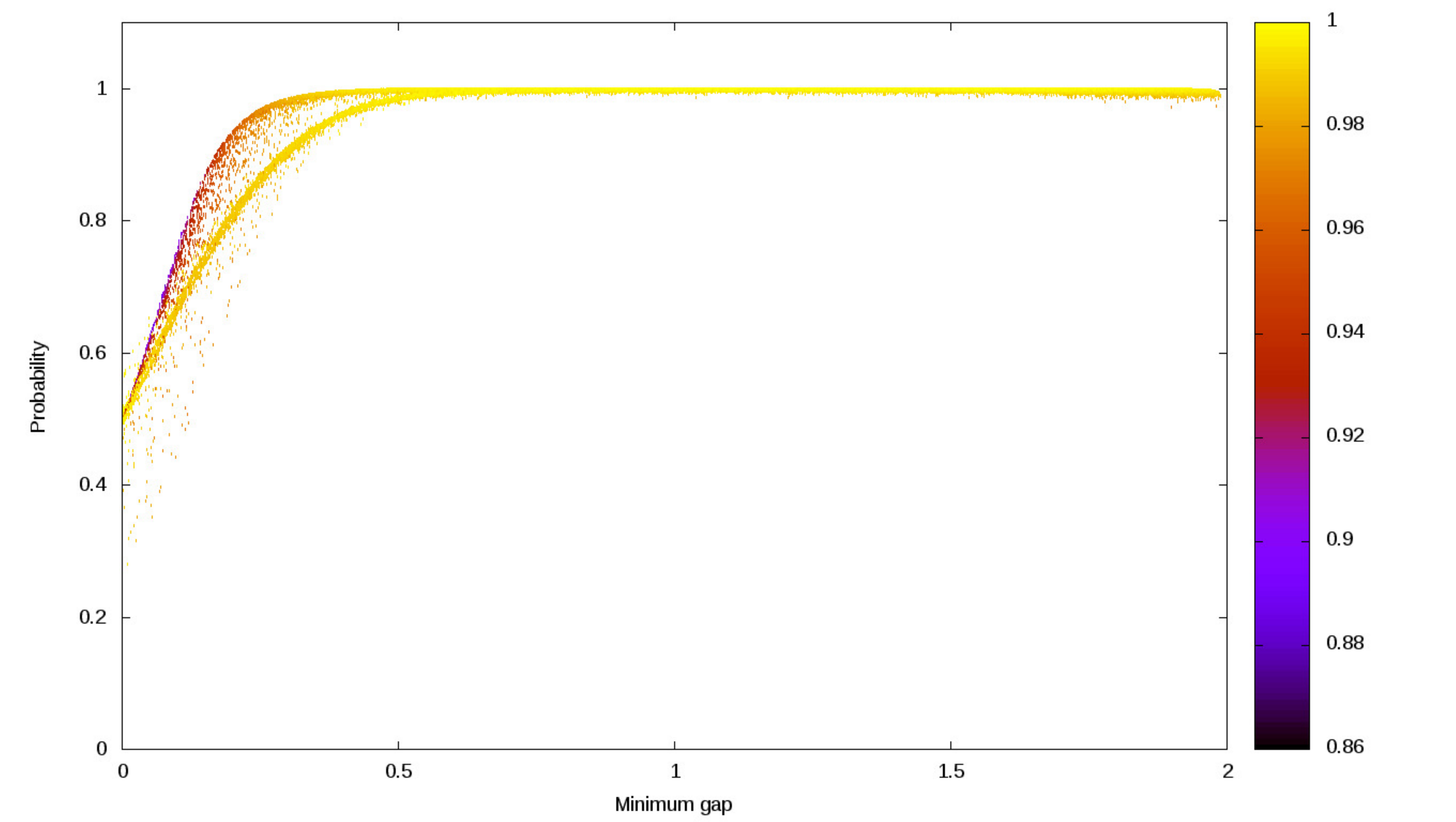}%
}%
\node (abscissa) at (3.25,-0.7) {\footnotesize Minimum gap ($\mingap$)};
\node (coodinate) [rotate=90] at (-1,2) {\footnotesize Probability ($P$)};
\node (figLabel) at (5.5,0.75) {\footnotesize $n=2$};

\node (0a) at (0.0,-0.15) {\scriptsize $0$};
\node (1a) at (1.625,-0.15) {\scriptsize $0.5$};
\node (2a) at (3.25,-0.15) {\scriptsize $1$};
\node (3a) at (4.875,-0.15) {\scriptsize $1.5$};
\node (4a) at (6.5,-0.15) {\scriptsize $2$};

\node (0c) [left] at (0,0.1) {\scriptsize $0$};
\node (0c) [left] at (0,0.88) {\scriptsize $0.2$};
\node (0c) [left] at (0,1.66) {\scriptsize $0.4$};
\node (0c) [left] at (0,2.44) {\scriptsize $0.6$};
\node (0c) [left] at (0,3.22) {\scriptsize $0.8$};
\node (0c) [left] at (0,4) {\scriptsize $1$};

\node (0b) [right] at (7,0.1) {\scriptsize $0.86$};
\node (0b) [right] at (7,2.2) {\scriptsize $0.93$};
\node (0b) [right] at (7,4.4) {\scriptsize $1$};
\end{tikzpicture}} 

\subfigure{
\begin{tikzpicture}[scale=0.85]
\pgftext[bottom,left]{%
\includegraphics[width=0.45\textwidth, trim = 40mm 20mm 50mm 0mm,clip]{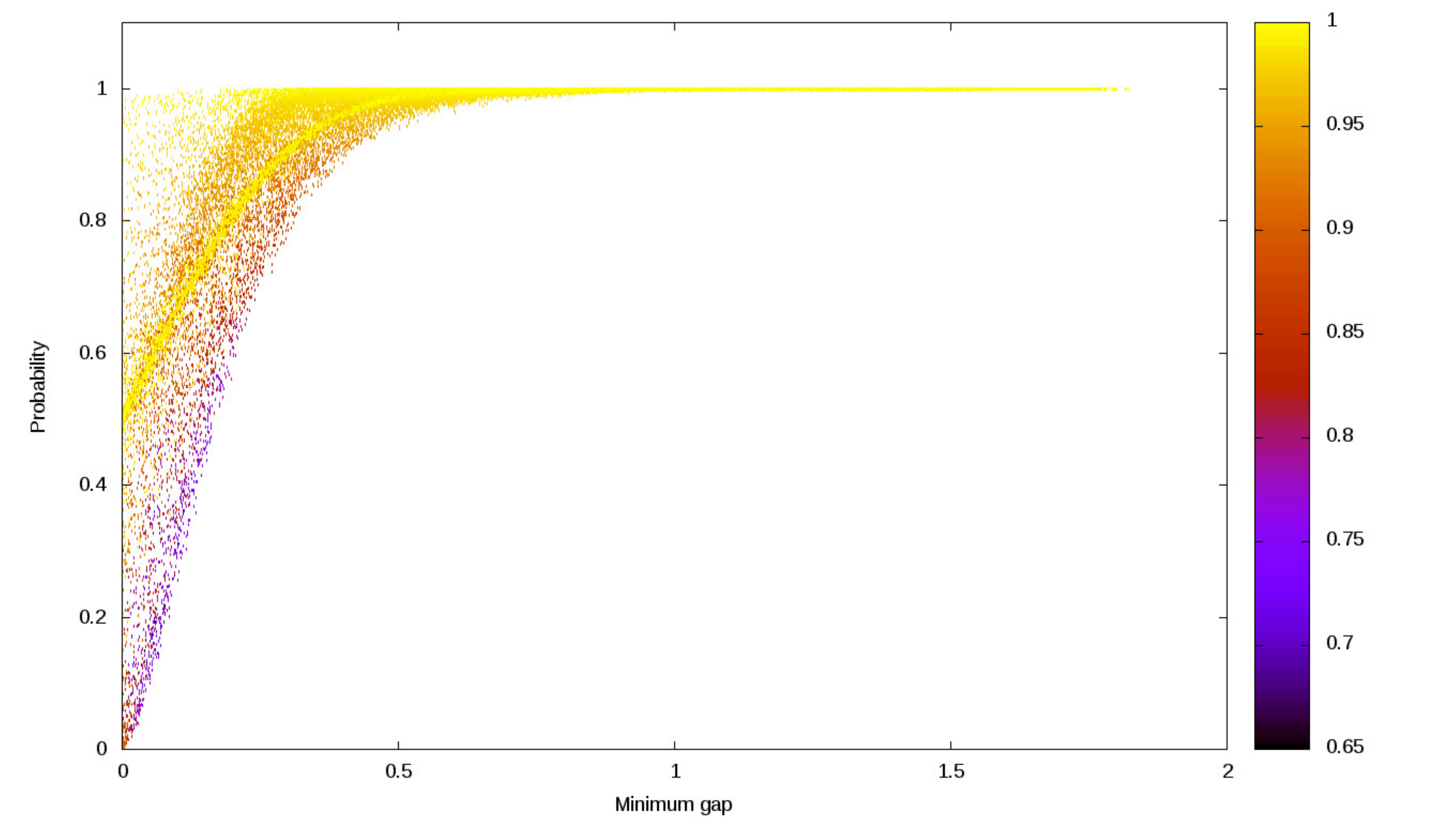}%
}%
\node (abscissa) at (3.25,-0.7) {\footnotesize Minimum gap ($\mingap$)};
\node (coodinate) [rotate=90] at (-1,2) {\footnotesize Probability ($P$)};
\node (figLabel) at (5.5,0.75) {\footnotesize $n=3$};

\node (0a) at (0.0,-0.15) {\scriptsize $0$};
\node (1a) at (1.625,-0.15) {\scriptsize $0.5$};
\node (2a) at (3.25,-0.15) {\scriptsize $1$};
\node (3a) at (4.875,-0.15) {\scriptsize $1.5$};
\node (4a) at (6.5,-0.15) {\scriptsize $2$};

\node (0c) [left] at (0,0.1) {\scriptsize $0$};
\node (0c) [left] at (0,0.88) {\scriptsize $0.2$};
\node (0c) [left] at (0,1.66) {\scriptsize $0.4$};
\node (0c) [left] at (0,2.44) {\scriptsize $0.6$};
\node (0c) [left] at (0,3.22) {\scriptsize $0.8$};
\node (0c) [left] at (0,4) {\scriptsize $1$};

\node (0b) [right] at (7,0.1) {\scriptsize $0.65$};
\node (0b) [right] at (7,2.2) {\scriptsize $0.825$};
\node (0b) [right] at (7,4.4) {\scriptsize $1$};
\end{tikzpicture}} 

\subfigure{
\begin{tikzpicture}[scale=0.85]
\pgftext[bottom,left]{%
\includegraphics[width=0.45\textwidth, trim = 40mm 20mm 50mm 0mm,clip]{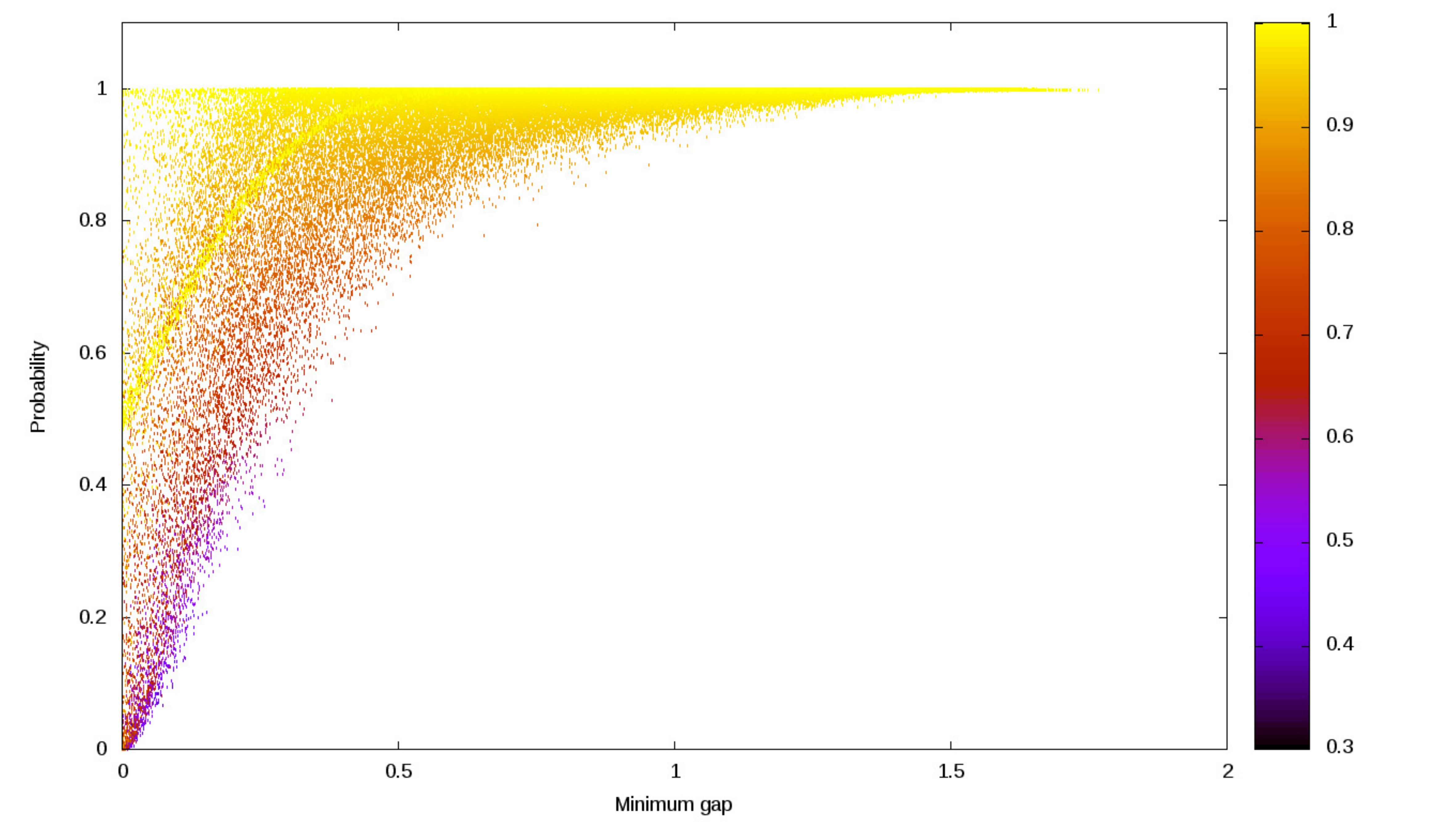}%
}%
\node (abscissa) at (3.25,-0.7) {\footnotesize Minimum gap ($\mingap$)};
\node (coodinate) [rotate=90] at (-1,2) {\footnotesize Probability ($P$)};
\node (figLabel) at (5.5,0.75) {\footnotesize $n=4$};

\node (0a) at (0.0,-0.15) {\scriptsize $0$};
\node (1a) at (1.625,-0.15) {\scriptsize $0.5$};
\node (2a) at (3.25,-0.15) {\scriptsize $1$};
\node (3a) at (4.875,-0.15) {\scriptsize $1.5$};
\node (4a) at (6.5,-0.15) {\scriptsize $2$};

\node (0c) [left] at (0,0.1) {\scriptsize $0$};
\node (0c) [left] at (0,0.88) {\scriptsize $0.2$};
\node (0c) [left] at (0,1.66) {\scriptsize $0.4$};
\node (0c) [left] at (0,2.44) {\scriptsize $0.6$};
\node (0c) [left] at (0,3.22) {\scriptsize $0.8$};
\node (0c) [left] at (0,4) {\scriptsize $1$};

\node (0b) [right] at (7,0.1) {\scriptsize $0.3$};
\node (0b) [right] at (7,2.2) {\scriptsize $0.65$};
\node (0b) [right] at (7,4.4) {\scriptsize $1$};
\end{tikzpicture}} 

\subfigure{
\begin{tikzpicture}[scale=0.85]
\pgftext[bottom,left]{%
\includegraphics[width=0.45\textwidth, trim = 40mm 20mm 50mm 0mm,clip]{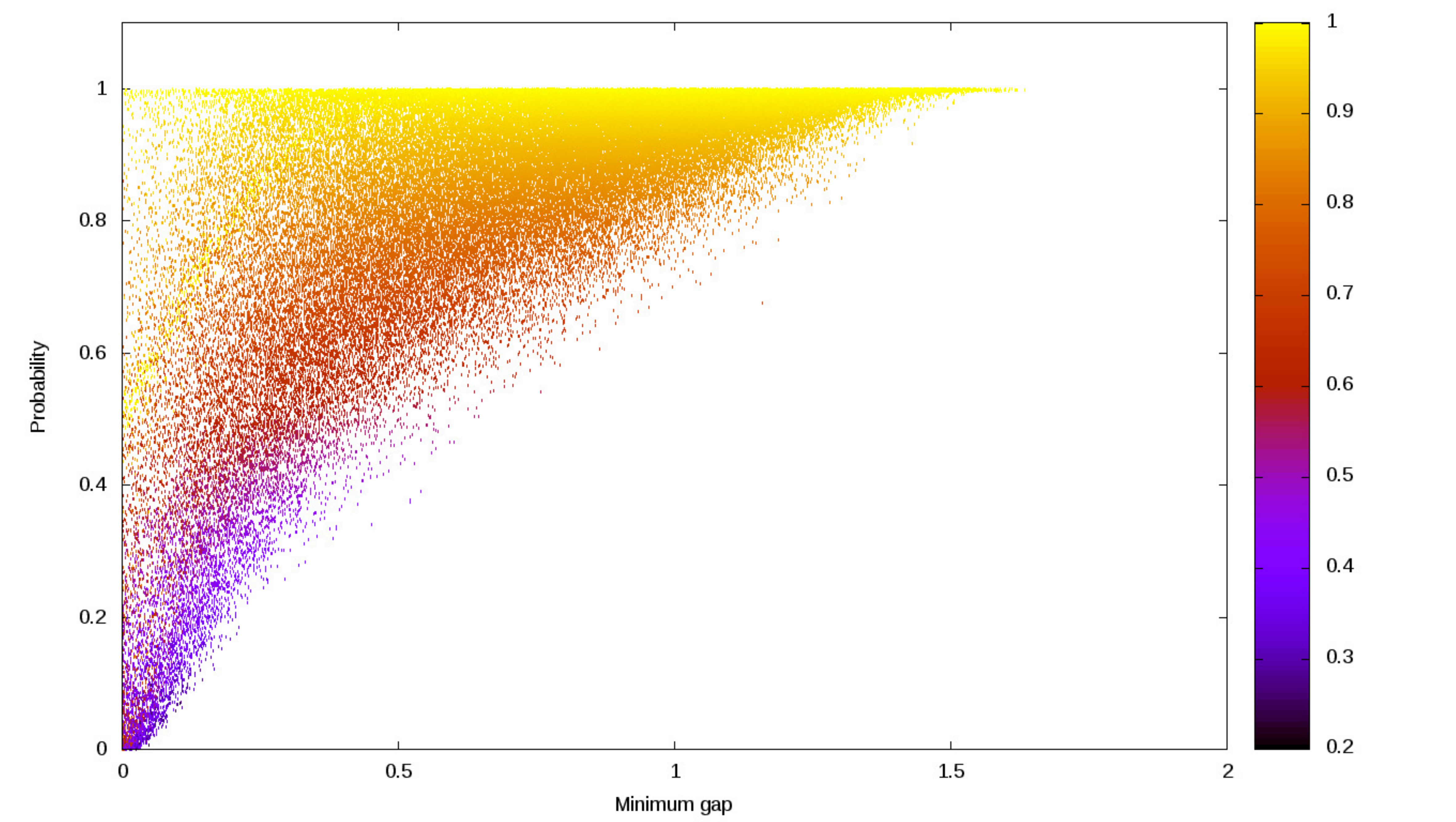}%
}%
\node (abscissa) at (3.25,-0.7) {\footnotesize Minimum gap ($\mingap$)};
\node (coodinate) [rotate=90] at (-1,2) {\footnotesize Probability ($P$)};
\node (figLabel) at (5.5,0.75) {\footnotesize $n=5$};

\node (0a) at (0.0,-0.15) {\scriptsize $0$};
\node (1a) at (1.625,-0.15) {\scriptsize $0.5$};
\node (2a) at (3.25,-0.15) {\scriptsize $1$};
\node (3a) at (4.875,-0.15) {\scriptsize $1.5$};
\node (4a) at (6.5,-0.15) {\scriptsize $2$};

\node (0c) [left] at (0,0.1) {\scriptsize $0$};
\node (0c) [left] at (0,0.88) {\scriptsize $0.2$};
\node (0c) [left] at (0,1.66) {\scriptsize $0.4$};
\node (0c) [left] at (0,2.44) {\scriptsize $0.6$};
\node (0c) [left] at (0,3.22) {\scriptsize $0.8$};
\node (0c) [left] at (0,4) {\scriptsize $1$};

\node (0b) [right] at (7,0.1) {\scriptsize $0.2$};
\node (0b) [right] at (7,2.2) {\scriptsize $0.6$};
\node (0b) [right] at (7,4.4) {\scriptsize $1$};
\end{tikzpicture}}
\end{center}
\caption{(Colour online) Distributions of success probability against $\mingap$ for two-, three-, four- and five-qubit systems over a set of $100,000$ random problem instances, with $T = 40$. The colouring of the points denotes $\delta$, the average overlap of the state vector with the instantaneous ground state (\ref{eq:overlap}).}
\label{fig:large_systems}
\end{figure}

\section{Larger systems}

We have shown that the relationship between the success probability and $\mingap$ is not a pure functional relationship for simple two-qubit systems. However, it is important to determine whether the interesting structure in this relationship remains in larger systems. To determine whether these densely-populated bands represent groups of problem instances that have followed similar evolution paths for the state vector (e.g.\ the system remaining mostly in the ground state, then being excited at a single avoided crossing), we calculated the average overlap with the ground state:
\begin{equation}
\delta = \int_0^1 \rmd s \bigl|\braket{0; s}{\psi(s)}\bigr|^2.
\label{eq:overlap}
\end{equation}
The points in figure \ref{fig:large_systems} are coloured with respect to this average overlap value, $\delta$, and we can see a smooth graduation across the figures, with the average overlap with the ground state increasing with the success probability. The exception to the smooth graduation of $\delta$ is the densely populated band where $\delta \approx 1$. This band must consist of cases with a degenerate or near-degenerate ground state at $s^{*} = 1$, as it includes cases which remain close to the instantaneous ground state throughout the majority of the evolution but have a low success probability. These results also lend credence to the idea that the structure is closely linked to the choice of Hamiltonian parameters.
We note that these distributions are reminiscent of the 2D projections of the higher-dimensional equilibrium surfaces seen in catastrophe theory \cite{saunders-1980}.
In this case success probability, $\mingap$ and $\delta$ are all internal variables of the system and not independent control parameters, so we are looking at a different situation to those usually studied in catastrophe theory. Identifying the nature of this surface and the dimensions of the phase space that it exists in is an important task, as it could have a major impact on adiabatic algorithm design. 

At this point we can  conjecture that the constraint originates from an adiabatic invariant of the Hamiltonian. $2^{2n}-1$ real parameters are required to specify the density matrix of $n$ qubits, reducing to $2^{n+1}-2$ for a pure state as discussed here.  The Pechukas-Yukawa approach to eigenvalue dynamics (see e.~g. Ref.~\cite{stoeckmann-1999}), which can be extended to density-matrix dynamics \cite{wilson-2011}, has at least $2^{n}$ adiabatic invariants, thus reducing the number of parameters required.   We find it strange that, to the best of our knowledge, there has been no research on adiabatic invariants of adiabatic quantum computers. We speculate that a  systematic investigation of adiabatic invariants of quantum computers --- especially adiabatic and approximately adiabatic computers --- could yield important information about their behaviour and have a major impact on  adiabatic algorithm design.

\section{Conclusion}

We have shown that the relationship between the success probability and the minimum ground state gap may not be as straightforward as is often assumed. There is a rich structure of distinct sharp edges and densely-populated bands in the distribution, particularly in smaller systems. A partial explanation has been proposed, whereby this is the projection of a higher-dimensional surface; identification of the parameters governing this surface will guide understanding of the set of problems amenable to adiabatic quantum computing. We do not propose a definitive explanation of the origin of this rich structure: this remains an open question. %However, we do suggest that there is some evidence that some of this structure could arise by our choice of Hamiltonians. 

\end{document}